\documentclass[letters,fleqn,usenatbib]{mnras}

\usepackage{natbib}
\usepackage{graphicx}
\usepackage{amssymb}
\usepackage{color}
\usepackage[dvipsnames]{xcolor}
\usepackage{caption}
\usepackage{lipsum,graphicx,multicol}
\usepackage{float}
\usepackage[fleqn]{amsmath}
\usepackage{subcaption}

\newcommand{\be}{\begin{equation}}
\newcommand{\ee}{\end{equation}}
\newcommand{\bq}{\begin{eqnarray}}
\newcommand{\eq}{\end{eqnarray}}

 \title {Estimating constraints on cosmological parameters via the canonical and the differential redshift drift  with SKA HI 21-cm observations}

 \author[]
  {
 	Jiangang Kang$^{1,2}$,
 	Guangyuan Song$^{3}$,
 	Tong Jie Zhang$^{1,2}$,\thanks{E-mail:  tjzhang@bnu.edu.cn}
 	Ming Zhu$^{4}$  
 	\\
 	$^{1}$Institute for Frontiers in Astronomy and Astrophysics, Beĳing Normal University, Beĳing 102206, China\\
 	$^{2}$	School of Physics and Astronomy, Beijing Normal University,  Beijing 100875, China\\
 	$^{3}$ Institute for Astronomical Science, Dezhou University, Dezhou 253023, People's Republic of China \\
 	$^{4}$National Astronomical Observatories, Chinese Academy of
 	Sciences,Beijing  100101, China\\}
 
 \date{Accepted XXX. Received YYY; in original form ZZZ}
 
\begin{document}
\maketitle
\label{firstpage} 
 
\begin{abstract}
		
 Redshift drift effect, an observational probe that indenpendent of cosmological models, presents unique applications in   specific cosmological epoch. By quantifying redshift drift signal , researchers can determine the rate of the Universe's accelerated expansion and impose constraints on  cosmological models and parameters. This study evaluates the precision in cosmological parameters estimation derived from this signal via  HI 21cm signal, that observed by the Square Kilometre Array (SKA) telescope, with spectral resolutions of 0.001 Hz and 0.002 Hz over an observational period of $\Delta T = 0.5$ year, utilizing two established techniques: the canonical redshift drift and the differential redshift drift method. The primary objective of this project is to ascertain the  rate of cosmic acceleration and establish a solid foundation for real-time cosmology. The results reveal that both the two methods impose highly precise constraints on cosmological parameters, with accuracy reaching  the level of  millimeter per second (mm/s) or better. However, the canonical method provides relatively less stringent  compared to the differential approach. Furthermore, when solely constraining the matter density parameter $\Omega_m$, the strategy can be adapted to the canonical method. Nonetheless, the differential  method exhibits clear advantages when simultaneously constraining  the matter density parameter $\Omega_m$ and the  equation of state of  dark energy. These findings validate SKA's capability in detecting redshift drift  and  refining  observational cosmology and  indicates the effect can offer superior diagnostic capabilities compared to other techniques, provided that appropriate observational equipment or sufficient observational time is employed.
\end{abstract}

\section{Introduction}\label{sec:intro}
Cosmological expansion causes temporal variations in the redshift of astronomical objects, known as the redshift drift effect or Sandage-Loeb effect, thoroughly studied in the literature \citep{Sandage,1998ApJ...499L.111L,Mcvittie,2015APh....62..195K}. Detecting this effect involves monitoring spectral line shifts in the Lyman-$\alpha$ forest or the  21 cm line from neutral hydrogen (HI) against the backgound bright sources, requiring exceptional observational precision, such as  (mm/s) per year or better \citep{1962ApJ...136..319S,1998ApJ...499L.111L,2007MNRAS.382.1623B,2022PDU....3701088L,2024RAA....24g5002K,Moresco_2022}. In addition, this probe serves as a unique tool for assessing the temporal evolution of cosmic acceleration, effectively functioning as a Cosmic Accelerometer \citep{Cooke_2019,2019BAAS...51g.137E,2022arXiv220305924C,2012PhRvD..86l3001M,2021MNRAS.508L..53E}. Presently, the universe is undergoing accelerated expansion driven by the repulsive dynamics of dark energy, characterized by its enigmatic physical properties; its equation of state (EoS) w(z) can be  investigated using observational data in defined contexts \citep{1998AJ....116.1009R,1999PhRvL..83..670P}. The quantification of redshift drift allows for a high-precision, independent differentiation of competing cosmological models and can resolve degeneracies when integrated with other observational data. Furthermore, it can characterize the properties of local cosmic structures and anisotropy by examining redshift drift fluctuations $\delta z$ across the sky using arbitrary photon paths \citep{2022PhRvD.106d3501K,2024arXiv240406242K,2023PhRvD.107f3544K,2021MNRAS.508L..53E}.

Regarding the Sandage-Loeb effect, which necessitates collecting redshift data from the same astronomical source over extended periods, often spanning many years or decades, it is crucial to note that this signal, being purely radial and cosmological, is inherently minuscule. Specifically, it manifests as approximately $\rm 10^{-10}$ and $\rm 2.9$ centimetre per second (cm/s) over about 12 years for redshift drift and velocity drift at redshift z=1, respectively \citep{2015aska.confE..27K}. The HI 21cm signal is widely recognized as a unique observational tool for quantifying the signal in ground-based experiments at redshift volumes $z<1$, whether through emission or absorption lines. Initially, this effect was detected using the Lyman-$\alpha$ forest observed with optical telescopes on Earth. However, this wavelength band becomes impractical for redshifts $z < 1.65$, particularly in the context of the  scientific investigations into the latest Universe's acceleration.

Damped Lyman alpha systems (DLAs), identified by their high-density neutral gas clouds comprising multiphase materials, are considered precursors as  the modern spiral galaxies and are characterized by HI column densities of $\rm N_{H I} \ge 2 \times 10^{20}/cm^2$\citep{1986ApJS...61..249W,2015A&A...575A..44G,2001A&A...369...42K,Kanekar_2001,Gupta_2013}. These dense or cold neutral hydrogen gas clouds absorb radiation from background luminous quasars, which is redshifted to 21 cm as it passes through these clouds, generating a distinctive absorption profile. This process yields vital insights into the host galaxy, including gas distribution, mass, temperature, kinematics, and star formation history, and enables the evaluation of cosmological evolution effects on the interstellar medium (ISM). Additionally, these absorption lines have recently been suggested as a feasible method for observing the Sandage-Loeb effect, which is independent of the redshift range of background absorbers and can be reliably detected by ground-based telescopes without atmospheric absorption or reflection. Similarly, 21 cm emissions from extragalactic sources at lower redshifts can also serve this purpose, although they are generally fainter compared to the former\citep{1988qsal.proc..297W,2001A&A...369...42K,Kanekar_2001,2015A&A...575A..44G,1998AJ....116...26L,2001A&A...373..394K,2015aska.confE.134M,2001A&A...369...42K,2023PASA...40...46K,Kanekar_2001,2024RAA....24g5002K}. Within the context of absorption line systems, Damped Lyman alpha (DLA) objects are generally classified as either intervening or associated, with three principal distinctions noted in \citep{2016MNRAS.462.4197C,2024RAA....24g5002K}. To obtain an accurate Sandage-Loeb signal and reduce the associated uncertainties, capturing the significant HI signals is imperative. This endeavor undoubtedly necessitates stable, high-caliber observational instruments with extremely high spectral resolution. Optimistically, the SKA telescope is poised to transform this field by cataloging vast samples of the H I 21cm lines with its advanced configurations. These catalogs can serve as candidate targets for detecting extragalactic emissions at redshifts up to 1 and the absorption lines of DLA systems extending to redshifts as high as 13\citep{2015aska.confE..27K,2004NewAR..48.1259K, 2019MNRAS.488.3607A,2020EPJC...80..304L,2023MNRAS.518.2853R,2021arXiv211012242M,2019arXiv190704495B,2015aska.confE.134M,2015aska.confE.167S,2015aska.confE..17A,marques2023watching,2024RAA....24g5002K,2022JApA...43..103D,2024MNRAS.tmp.2497E,2024arXiv240806626Y}. Simultaneously, the European Extremely Large Telescope (ELT) is set to observe the Sandage-Loeb signal at redshifts ranging from 2 to 5 via the Lyman-alpha forest, systematically surveying the southern celestial hemisphere over a 20-year timeframe \citep{2024arXiv240806626Y, 2021arXiv211012242M, Liske_2008, 2013arXiv1310.3163M, 2021Msngr.182...27M}. Furthermore, existing observatories such as CHIME and FAST possess the potential to accomplish analogous goals at intermediate and lower redshifts, respectively, provided that their spectral line precision undergoes enhancement \citep{Yu_2014,Newburgh_2014,Bandura_2014,NAN_2011,8331324,2019SCPMA..6259502J,Jiao_2020,2023A&A...675A..40H}.
 
In this investigation, we explore the constraints on the precision of cosmological parameters $\rm\sigma_{p}$ (where p represents an individual parameter) at the level of less than  cm/s/year by detecting the redshift drift effect with datasets of the spectral resolution of 0.001 Hz and 0.002 Hz with SKA, by the observing period of $\Delta T$ = 0.5 year, under the Chevallier-Polarski-Linder (CPL) paradigm \citep{CPL1,CPL2}. The CPL parameterization competently captures the evolving nature of dark energy over cosmic epochs \citep{2024arXiv241209409O,2025arXiv250314738D} and provides superior fits to observational data compared to the $\rm\Lambda$CDM and wCDM models, as evidenced by recent datasets from large-scale survey projects \citep{2017NatAs...1..627Z}. The CPL model's responsiveness to new data enhances its efficacy for precise cosmological forecasts \citep{2024arXiv241215124G}.
The precision of the parameter is characterized by the partial derivative of the velocity drift concerning the parameter, $\rm \partial (\Delta v)/\partial p$, where $\rm \Delta v$ denotes the velocity drift and p represents one of the model parameters $(h,\Omega_m,w_0,w_a)$. This analysis is based on the data of redshift drift signal acquired at spectral resolution channels of 0.001 Hz and 0.002 Hz on SKA over an experimental duration of $\Delta T = 0.5$ year. Two established techniques are employed to detect the redshift drift signal: the  canonical  redshift drift method for measuring drift relative to the current epoch \citep{Eikenberry}, and the differential redshift drift technique \citep{Cooke}, which assesses two distinct non-zero redshifts by concurrently measuring the redshift of two objects, specifically the intervening and reference object, with $\rm z_i < z_r$. This study presumes a spatially flat cosmological model throughout and is structured as follows: Section \ref{sec:intro} provides an overview of the detection of redshift drift and cosmology, including details on the HI 21cm signal and SKA observations; Section \ref{sec:mo} explains the background of the redshift drift  effect and the methodologies used for parameter precision estimation; Section \ref{sec:sys} investigates potential systematic effects and the criteria for source selection; Section \ref{sec:meas} presents results from the constraints derived from  the data of redshift drift mesurement using the HI 21cm  signals with SKA; finally, Section \ref{sec:sum} offers a comprehensive summary of the main outcomes and conclusions of this research.
\section{redshift drift model and methodology}\label{sec:mo}
\begin{figure}
 	\begin{center}
 		\includegraphics[width=.5\textwidth]{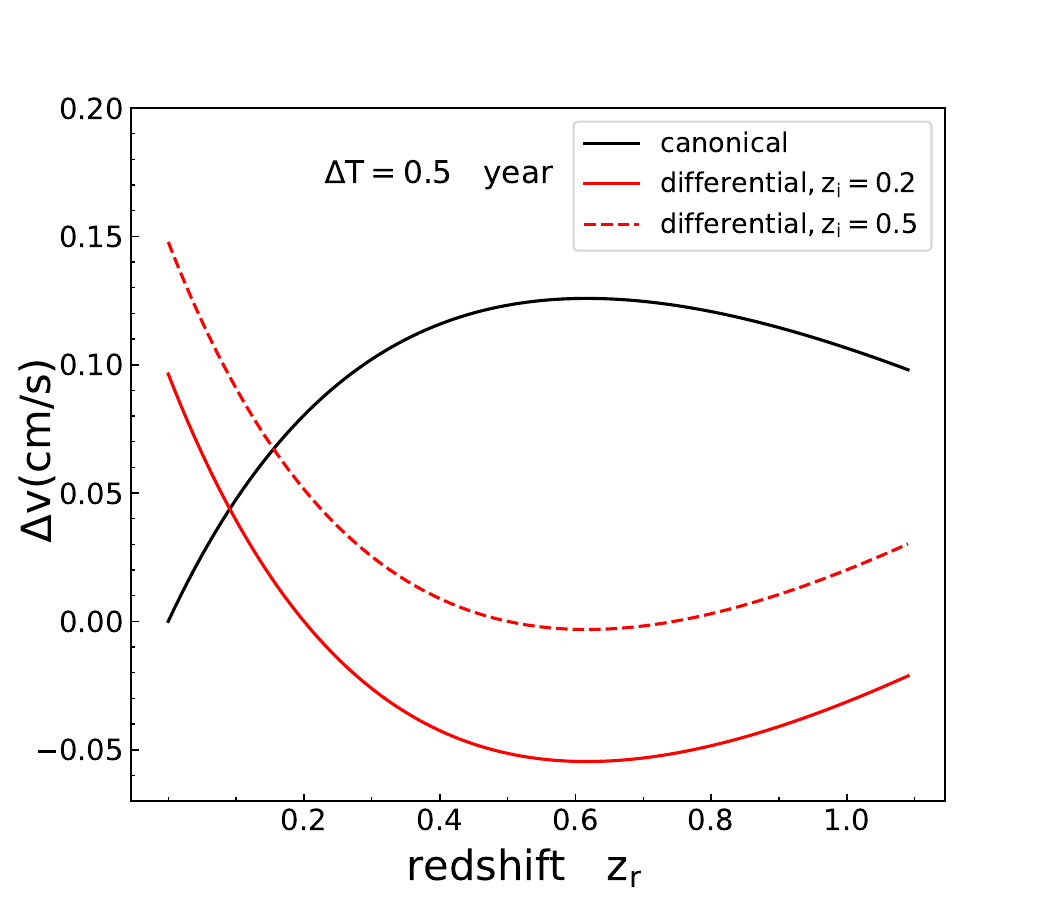}
 		
 	\end{center}
 	\caption{The theoretical amplitude of the canonical and differential redshift drift methodologies as a function of the reference redshift $\rm z_r$ over the observing period $\Delta T$ = 0.5 year. The black solid line illustrates the canonical redshift drift values, while the red solid and dashed lines represent the differential redshift drift values at intervening redshifts $\rm z_i$ = 0.2 and 0.5, respectively.}
 	\label{figs:sd} 
\end{figure}

The redshift variation $\Delta z$ of an astronomical object results from the continuous expansion of the Universe over a time interval $\rm \Delta t$ \citep{Sandage}. When this effect is measured relative to the current epoch (redshift z=0), it is referred to as the canonical redshift drift method and can be mathematically represented as follows:
\be\label{eqs:1}
\frac{\Delta z}{\Delta t}=H_0 \left[1+z-E(z)\right]\,,
\ee
Typically, the phenomenon is measured via the spectroscopic shifts in radial velocity,
\be\label{eqs:11}
\Delta v=\frac{c\Delta z}{1+z}=(cH_0\Delta t)\left[1-\frac{E(z)}{1+z}\right]\,,
\ee
Within this context, $E(z)=H(z)/H_0$ represents the dimensionless Hubble parameter, where $H_0$ denotes the Hubble constant. Furthermore, the differential redshift drift for objects with non-zero redshift along a common line of sight can be expressed as follows\citep{Cooke,2021MNRAS.508L..53E}:

\be\label{eqs:2}
\Delta v_{ir}=(cH_0\Delta t)\left[\frac{E(z_r)}{1+z_r}-\frac{E(z_i)}{1+z_i}\right]\,,
\ee
In this context, $z_r$ and $z_i$ represent the redshifts of the reference and intervening sources, respectively, under the condition that $z_i < z_r$. It appears more advantageous to employ the differential redshift drift technique, as it slightly improves the detectability of the drift signal compared to the traditional method, as illustrated in figure \ref{figs:sd}. This suggests that the magnitude of the former is marginally larger than that of the latter. Furthermore, as the intervening redshift $z_i$ decreases, the amplitude of the redshift drift signal $\Delta v$ increases. Over a period of $\rm\Delta T$ = 0.5 year, the amplitude of $\Delta v$ ranges from 0 to 0.15 cm/s when using the differential redshift drift technique, whereas it ranges from 0 to 0.12 cm/s with the traditional method. The CPL parameterization, which includes four parameters $(H_0, \Omega_m, w_0, w_a)$, is adopted as the standard model in this study, with its corresponding Hubble parameter equation expressed as follows:
\be\label{cpl}
E^2(z)=\Omega_m(1+z)^3+\Omega_\phi(1+z)^{3(1+w_0+w_a)} e^{-3w_az/(1+z)}\,.
\ee
The dimensionless redshift drift can subsequently be accurately measured\citep{2021MNRAS.508L..53E,2019MNRAS.488.3607A}:
\be\label{eqs:s0}
S_z=\frac{1}{H_{100}}\frac{\Delta z}{\Delta t}=h\left[1+z-E(z)\right]\,,
\ee
Assuming \( H_0 = hH_{100} \) where \( H_{100} = 100 \) km/s/Mpc, the corresponding measurable spectroscopic velocity shift can be formulated as:
\be\label{eqs:s00}
S_v={\Delta v}=kh\left[1-\frac{E(z)}{1+z}\right]\,,
\ee
which is expressed in cm/s, the constant $ k$ is given by $\ k=cH_{100}\Delta T$, with $\Delta T$ fixed at 0.5 year, taking $k$ being 1.532 cm/s. Under the assumptions of the fiducial flat CPL model, the observational precision of the parameters, represented as $\rm \partial S_{(z,v)}/ \partial p_i$, is described by equations \ref{eqs:s1}-\ref{eqs:s4}:
\be \label{eqs:s1}
\frac{\partial S_z}{\partial h}=1+z-E(z),
\ee
\be \label{eqs:s2}
\frac{\partial S_z}{\partial\Omega_m}=-\frac{h(1+z)^3}{2E(z)}\left[1-(1+z)^{3(w_0+w_a)}\exp{\left[\frac{-3w_az}{1+z}\right]}\right],
\ee
\be \label{eqs:s3}
\frac{\partial S_z}{\partial w_0}=-\frac{3h(1-\Omega_m)}{2E(z)}(1+z)^{3(1+w_0+w_a)}\ln{(1+z)}\exp{\left[\frac{-3w_az}{1+z}\right]},
\ee

\be \label{eqs:s4}
\begin{aligned}
\frac{\partial S_z}{\partial w_a}=&-\frac{3h(1-\Omega_m)}{2E(z)}(1+z)^{3(1+w_0+w_a)}\left[\ln{(1+z)}-\frac{z}{1+z}\right]\\
 & \times  \exp{\left[\frac{-3w_az}{1+z}\right]}.
\end{aligned}
\ee
 
\begin{figure}
	\begin{center}
	\includegraphics[width=.56\textwidth]{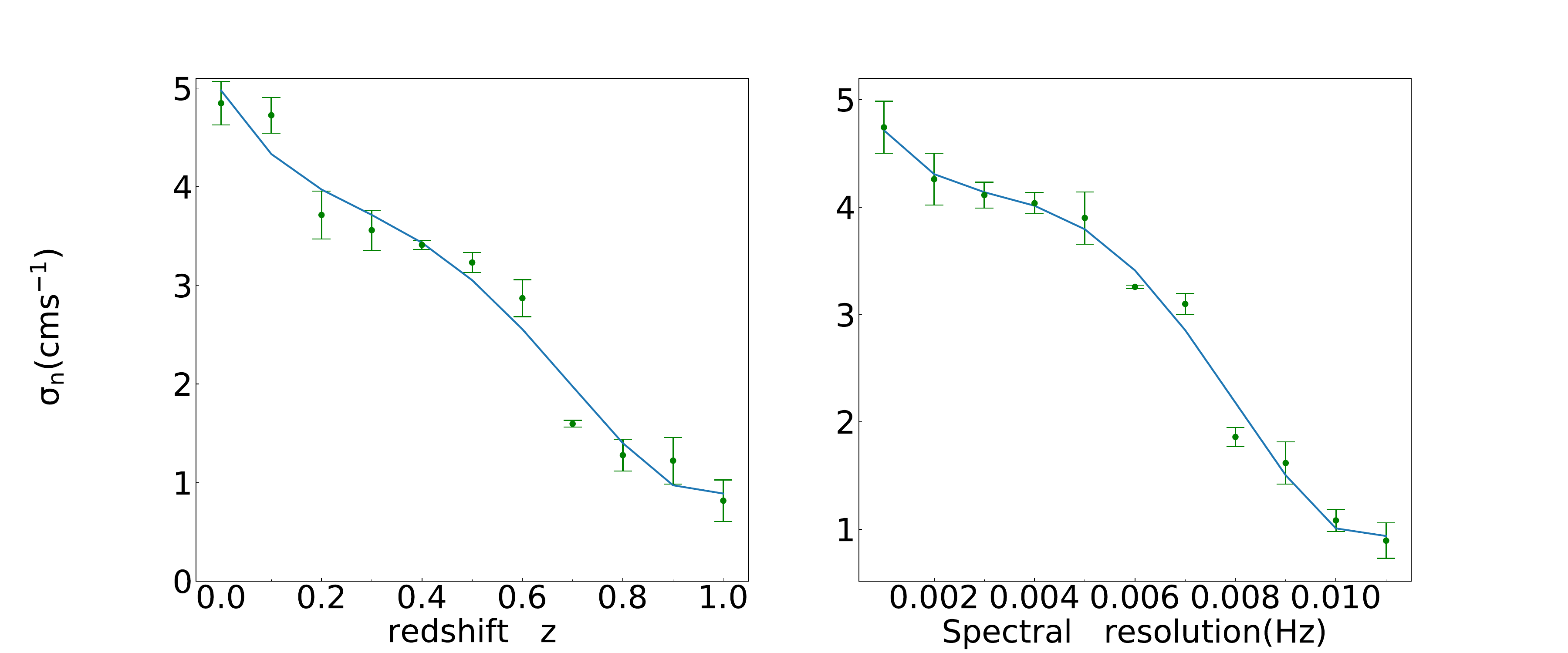}
	 
	\end{center}
	\caption{The relationship of the normalized constant $\sigma_{n}$ with the spectral resolution and redshift is illustrated, with the green error bars indicating the actual measurements and the blue solid line representing the fitted values.}
	\label{figs:1} 

\end{figure}

The study targets the examination of the precision constraints and uncertainties of cosmological parameters using the redshift drift measurements obtained by SKA. By employing spectral resolutions of 0.001 Hz and 0.002 Hz, it is demonstrated that the spectroscopic velocity uncertainty, denoted as $\rm\sigma_{v}$, is contingent upon the total number of HI 21cm signals (N), encompassing both emission and absorption lines, as well as the observational time span $\Delta T$. This relationship can be precisely modeled as follows:
\be\label{eqs:sv1}
\rm \sigma_v=\sigma_n N^{-1/2}(1+z)^{\lambda}\rm \Delta T^{-1/2}   \quad [ cm/s],
\ee
The parameter $\rm\sigma_{n}$ represents a normalization constant. According to the estimate that there are $N=10^7$ objects per 0.1 redshift interval up to a redshift of 1, within an observational period of $\Delta T = 0.5$ year, utilizing spectral resolutions of 0.001 Hz and 0.002 Hz. The parameter $\lambda$ is assigned values of 1.09 for the 0.001 Hz data and 1.52 for the 0.002 Hz data. The SKA is designed to detect redshift drift signal with spectral resolutions spanning from 0.001 Hz to 0.01 Hz \citep{2015aska.confE..27K,2009ApJ...703.1890O,2019arXiv190704495B,2015MNRAS.450.2251Y}. For the experiment at redshift $z \le 1$, the preferred spectral resolution should not exceed 0.002 Hz, as indicated by the evaluation of equations \ref{eqs:s0}-\ref{eqs:s00}.   Figure \ref{figs:1} illustrates the correlation of the normalization constant $\sigma_{n}$ with the spectral resolution and redshift. The green error bars denote the observed data, while the blue solid line represents the fitted curve. As shown in Figure \ref{figs:1}, $\rm\sigma_{n}$ exhibits a linear decline with respect to both redshift and spectral resolution. Lower redshifts or higher spectral resolutions correspond to larger values. Except when the redshift $z$ exceeds 0.8 or the spectral resolution is above 0.01 Hz, these values remain stable at approximately 1 cm/s. Consequently, the ultra-high spectral resolution and a substantial number of sampled objects with adequate signal-to-noise ratios (S/N) during the SKA observational era will ensure the experiment's reliability.

\section{Systematic effects and Target selection}\label{sec:sys}

Systematic biases include the baseline stability and frequency calibration accuracy, the angular displacement of the absorbing gas relative to the background sources, fluctuations in the size, flux, and spectral properties of the illumination source, observer kinematics, peculiar velocities and accelerations. The required calibration precision is dependent on the local oscillator, and current radio observatories possess the capability to maintain this level of accuracy\citep{2020MNRAS.492.2044C}. Despite the experiment being conducted within a highly non-inertial reference frame characterized with multiple accelerations and rotations, the precision achievable in the era of SKA will be sufficiently refined, even surpassing the necessary requirements for the detection of redshift drift effect  \citep{2015aska.confE..27K,2020MNRAS.492.2044C,2015APh....62..195K,Moresco_2022,2019MNRAS.488.3607A}. Moreover, it is widely recognized that the  sign of  $\dot{z}$, deviates from zero at both low and high redshifts. Additionally, the directions of gravitational accelerations exhibit random behavior and are centered around a null mean. Therefore, the aggregate effect of peculiar accelerations can be diluted by enlarging the HI 21cm sample sizes across various sky regions, enhancing integration time, and prolonging the experimental  duration\citep{Moresco_2022}. In terms of the plausibility of conducting experiments with  SKA telescope \citep{2015aska.confE..27K}, for extragalactic sources with redshifts surpassing 0.2, the peculiar acceleration in redshift space will be diminished to $10^{-14}$, a factor of 10 less than  cosmological signal at the percent-level, thereby rendering them negligible in  observational studies\citep{Liske_2008,2019BAAS...51g.137E,2012PhR...521...95Q,2024arXiv240909977B,2024arXiv240406242K,2023JCAP...11..093B}.

Beyond the frequency stability and gravitational influences on the HI 21cm signals, a manifold of technical parameters are demanded to the comprehensive scrutiny. These impacts can be broadly categorized into three primary domains: Firstly, the observations will be executed within an ultra-stable inertial reference frame, ensuring no temporal or spatial discrepancies with respect to the targets. Astrometric precision, temporal standards, and pointing accuracy must be calibrated and refined to within 1 arcsecond. These inherent interferences will be systematically mitigated as data from spectral channels are ingested into the processing pipelines. Secondly, the precision timing capabilities inherent in pulsar observations offer an impartial methodology for evaluating the long-term stability of the global SKA system. By precisely measuring the time-of-arrival of pulses from multiple pulsars, it becomes feasible to mitigate systematic biases. Moreover, an extensive array of correlator channels will be utilized to thoroughly capture the neutral hydrogen emissions from  extragalaxies. Achieving the measurement of redshift drift necessitates between $10^{8}$ $\sim$ $10^{9}$ correlator channels, a complexity that can be efficiently addressed through precise spectral window configuration and the standardization of frequency during data acquisition\citep{2015aska.confE..27K,2020PASA...37....7S}. 

Undoubtedly, the exceptional sensitivity of the SKA enables the acquisition of ample radiation emitted by targeted objects, thereby maintaining acceptable error margins. For HI 21cm emission, galaxies exhibiting a singular Gaussian profile are deemed optimal candidates, provided their peak flux density surpasses 100 mJy and their signal-to-noise ratio (S/N) is at least 100. This preference arises due to the superior baseline stability and outstanding sensitivity of the SKA apparatus. Conversely, DLA systems characterized by multiple Gaussian components yield reduced uncertainties for Sandage-Loeb signal, thereby establishing intervening-DLA systems as the primary candidates\citep{2020MNRAS.492.2044C,2023arXiv230204365C,Darling_2012,2020EPJC...80..304L,2015aska.confE.167S,2015aska.confE..17A,2022PASA...39...10A,1988qsal.proc..297W,2001A&A...369...42K,2024MNRAS.tmp.2497E,2024arXiv240806626Y,2024MNRAS.tmp.2497E,2015APh....62..195K,2023MNRAS.518.2853R,2004NewAR..48.1259K,2020PASA...37....7S,2015aska.confE.134M,2024RAA....24g5002K,2015MNRAS.450.2251Y,2022JApA...43..103D}. 
\section{measurement versus constraint}\label{sec:meas}

\begin{figure*}
\begin{center}
\includegraphics[width=\columnwidth,keepaspectratio]{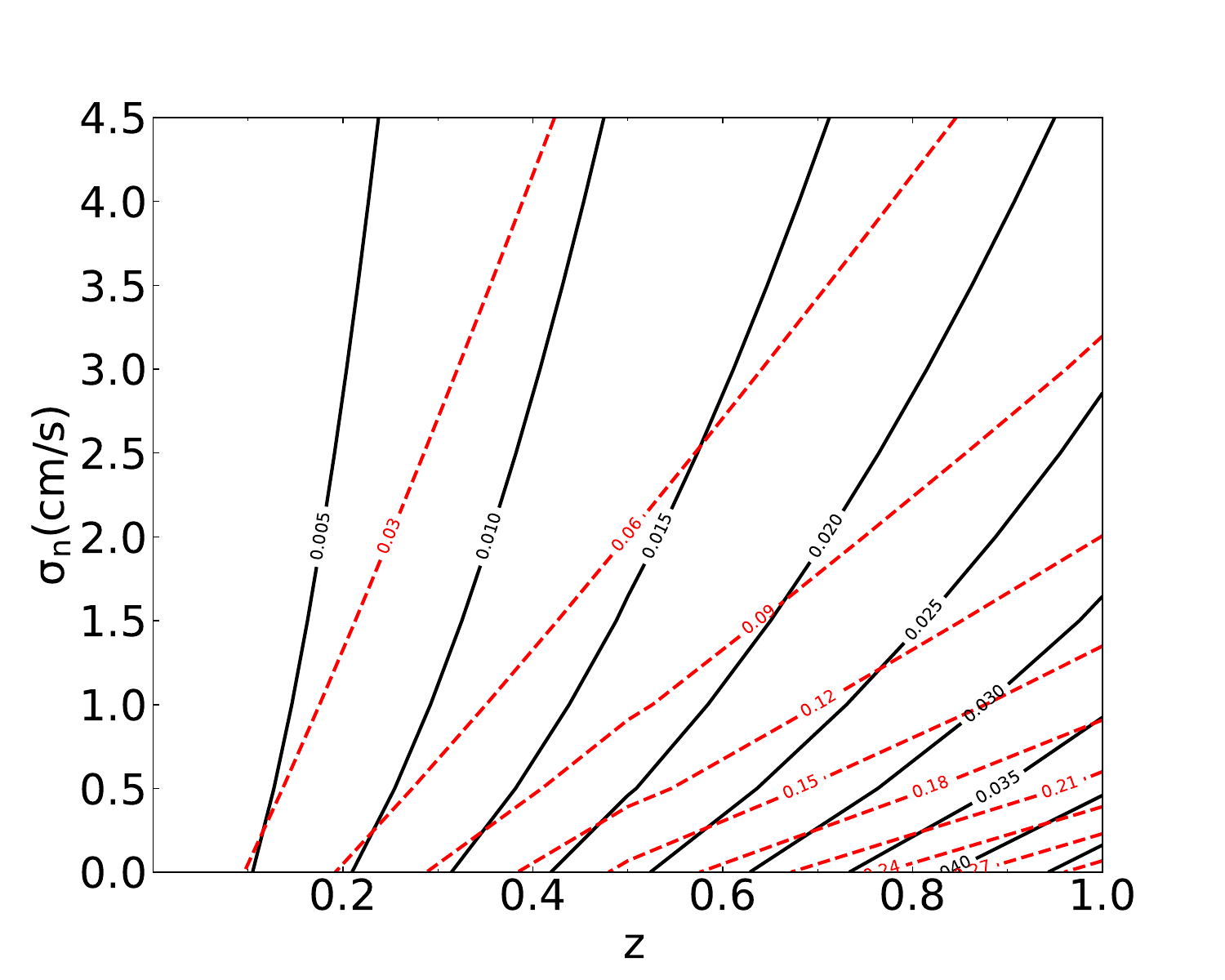}
\includegraphics[width=\columnwidth,keepaspectratio]{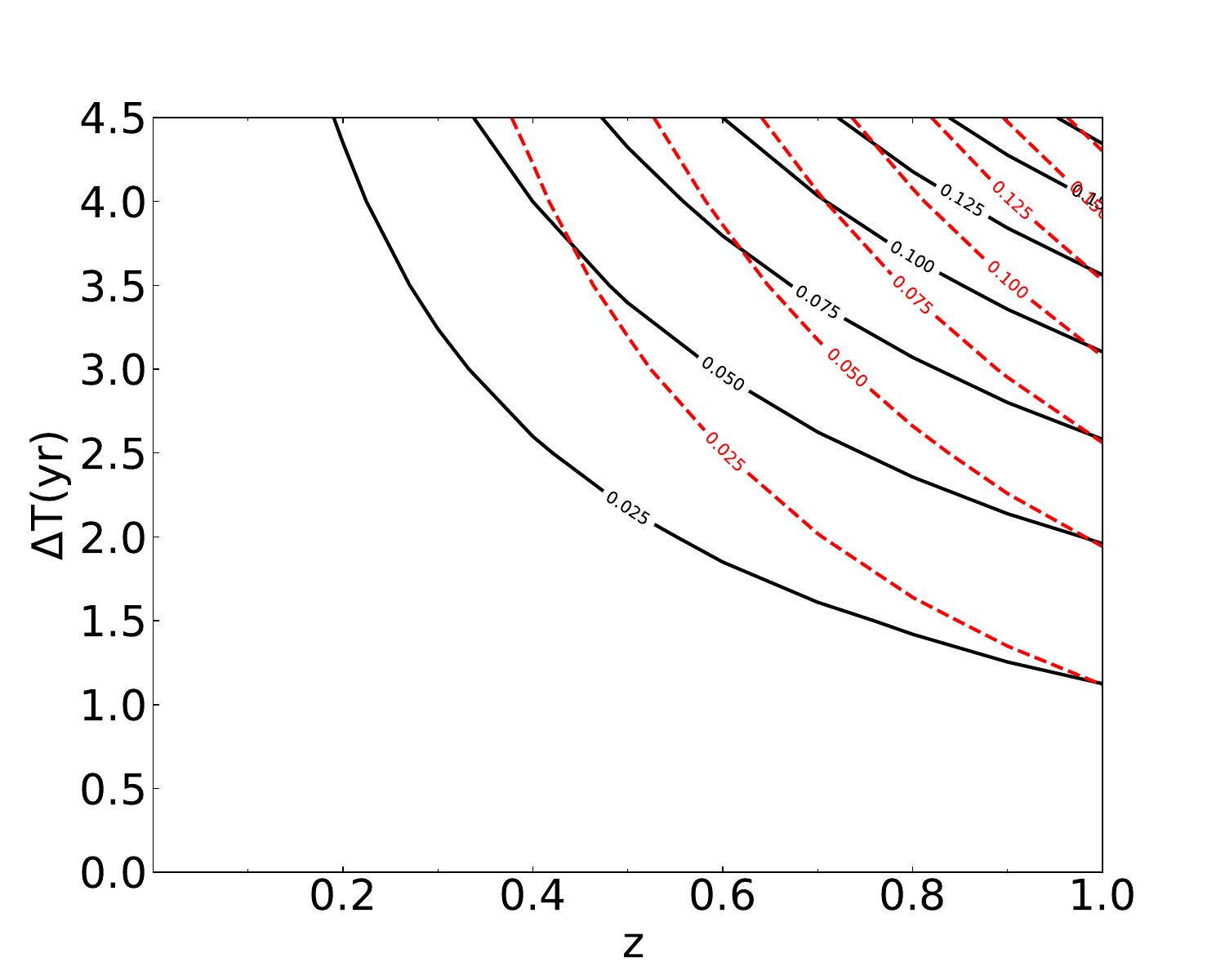}
\includegraphics[width=\columnwidth,keepaspectratio]{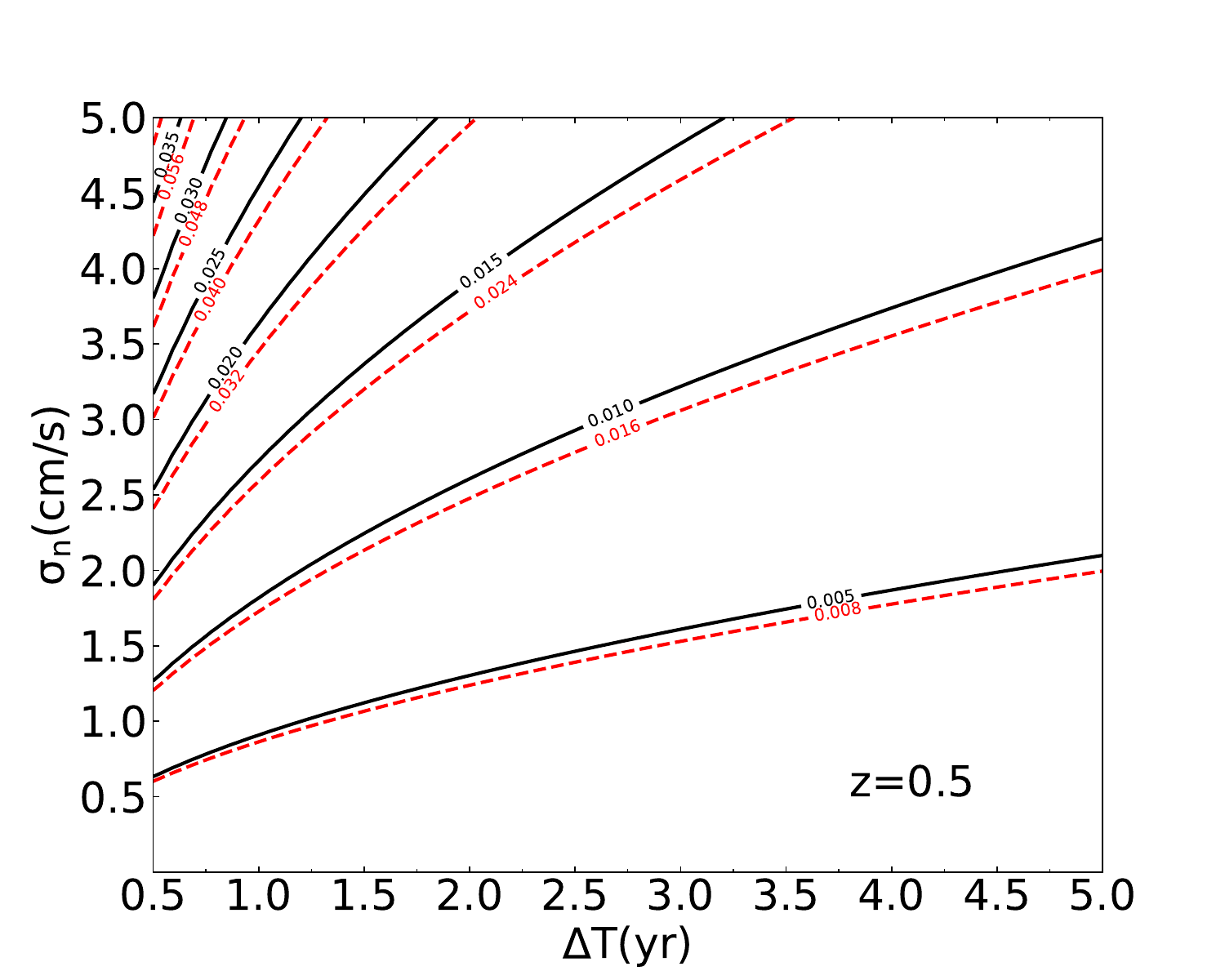}
\includegraphics[width=\columnwidth,keepaspectratio]{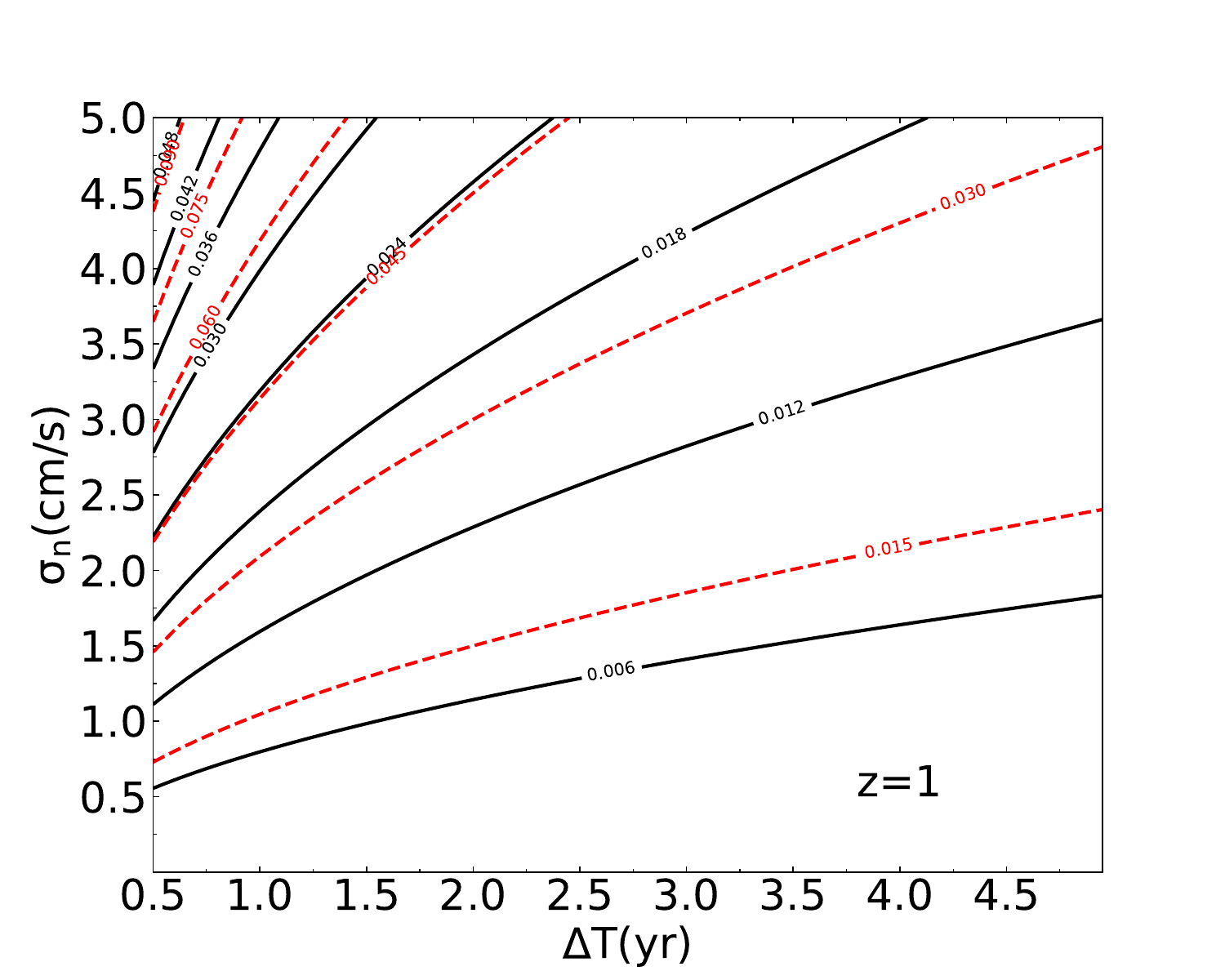}
\end{center}
\caption{The uncertainty in spectroscopic velocity drift ($\sigma_v$) is depicted as a function of redshift ($z$), observational period ($\Delta T$), and normalization constant ($\sigma_n$) in equation \ref{eqs:sv1}. In each plot, the black solid and red dotted contours illustrate the measured outcomes for spectral resolutions of 0.001 Hz and 0.002 Hz, respectively. The upper two panels demonstrate how the relationship ($\sigma_v$) varies with the normalization constant and observational period within the redshift domain, whereas the lower two panels fix the redshifts at $z=0.5$ and $z=1$, respectively, with  $\Delta T$ ranging from 0.5 to 5 years.}
 \label{figs:sv1}
\end{figure*}

  The measurement of redshift drift effect is accomplished utilizing the spectral resolutions of 0.001 Hz and 0.002 Hz on the SKA telescope. The investigation initially assesses the uncertainties in the spectroscopic velocity $\rm\sigma_{v}$, which are collectively influenced by the normalization constant $\sigma_n$, the observation period $\Delta T$, and the redshift z. These assessments are derived from equation \ref{eqs:sv1}, and the results are illustrated in figure \ref{figs:sv1}, demonstrating that $\rm\sigma_{v}$ is directly proportional to both the redshift z and $\sigma_{n}$, yet inversely proportional to the observation period $\Delta T$  that resulted from the canonical redshift drift. The contour numbers of each plot in figure \ref{figs:sv1} indicate the magnitude level of $\rm \sigma_v$ as a function of the quantities on the x-axis and y-axis, here N denotes the total number of sources, approximately $\rm 10^8$, with $\lambda$ fixed at 1.09 for 0.001Hz data and 1.52 for 0.002Hz data. In the top-left plot of figure \ref{figs:sv1}, $\rm\sigma_v$ ranges from 0.005 to 0.045 cm/s for the 0.001Hz case and 0.03 to 0.3 cm/s for the 0.002Hz case, while the observation period $\Delta T$ is fixed at 0.5 year.  In the top-right plot of figure \ref{figs:sv1}, $\rm\sigma_v$ spans from 0.025 to 0.15 cm/s for 0.001Hz and 0.025 to 0.2 cm/s for 0.002Hz, where the normalization constant $\sigma_n$ is fixed at 3 cm/s. The bottom-left plot of figure \ref{figs:sv1} shows changes in $\rm\sigma_v$ from 0.005 to 0.035 cm/s for 0.001Hz and from 0.008 to 0.056 cm/s for 0.002Hz at redshift z = 0.5. In the bottom-right plot of figure \ref{figs:sv1}, $\rm\sigma_v$ ranges from 0.006 to 0.048 cm/s for 0.001Hz and from 0.015 to 0.09 cm/s at redshift z = 1.

 As a result, the values of $\rm\sigma_{v}$ range from approximately 0.005 to 0.15 cm/s for the 0.001 Hz scenario, and from 0.02 to 0.3 cm/s for the 0.002 Hz scenario, while maintaining errors below 1 cm/s for redshifts up to z = 1 when employing the canonical redshift drift method. This demonstrates that these datasets possess adequate precision for the signal detection. Additionally, the findings imply that prolonging the observation period can substantially amplify the redshift drift signal's magnitude and markedly diminish systematic uncertainties, provided that the observed redshift range is not excessively extensive.
   
\begin{figure*}
\begin{center}
\includegraphics[width=.33\textwidth]{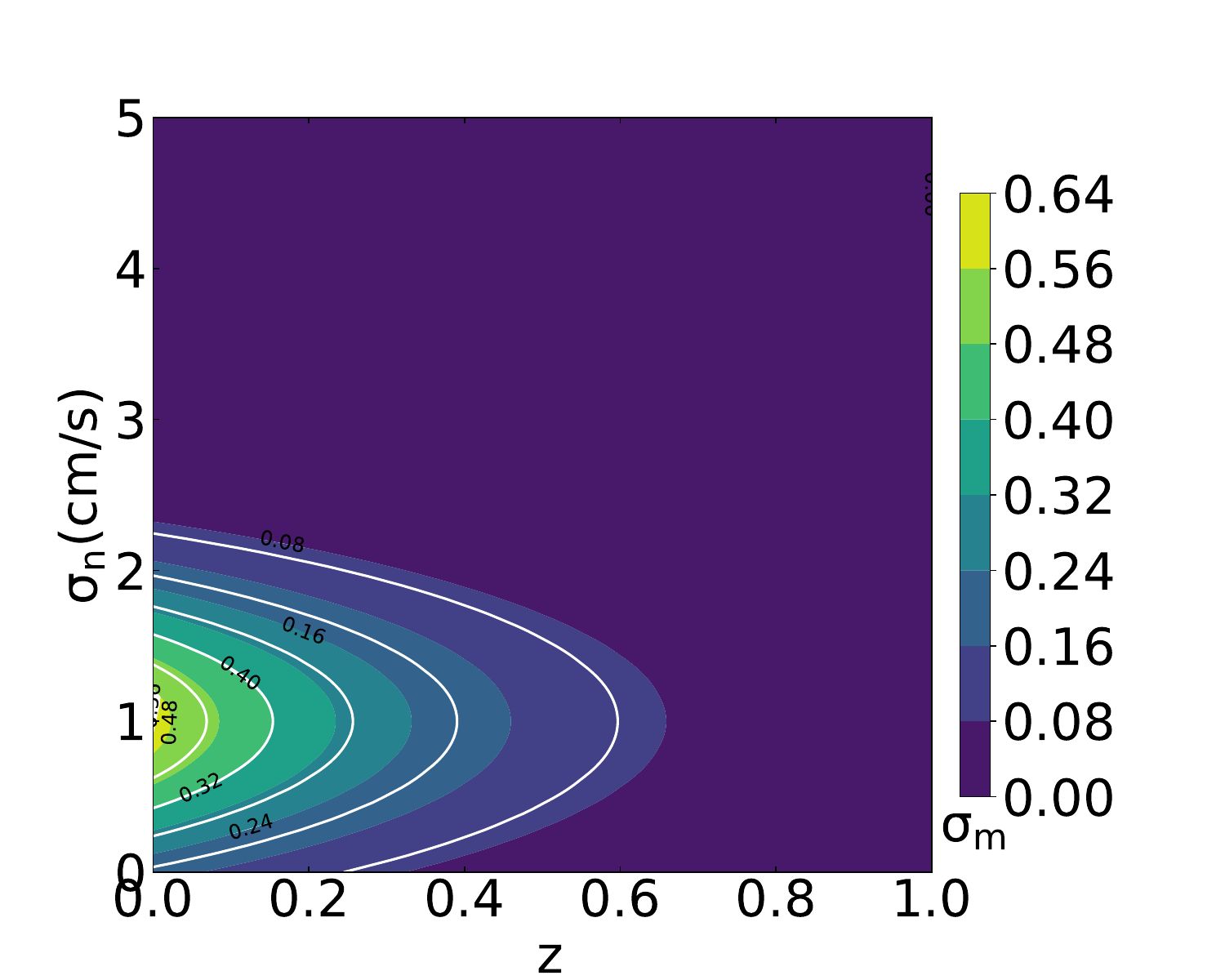}
 \includegraphics[width=.323\textwidth]{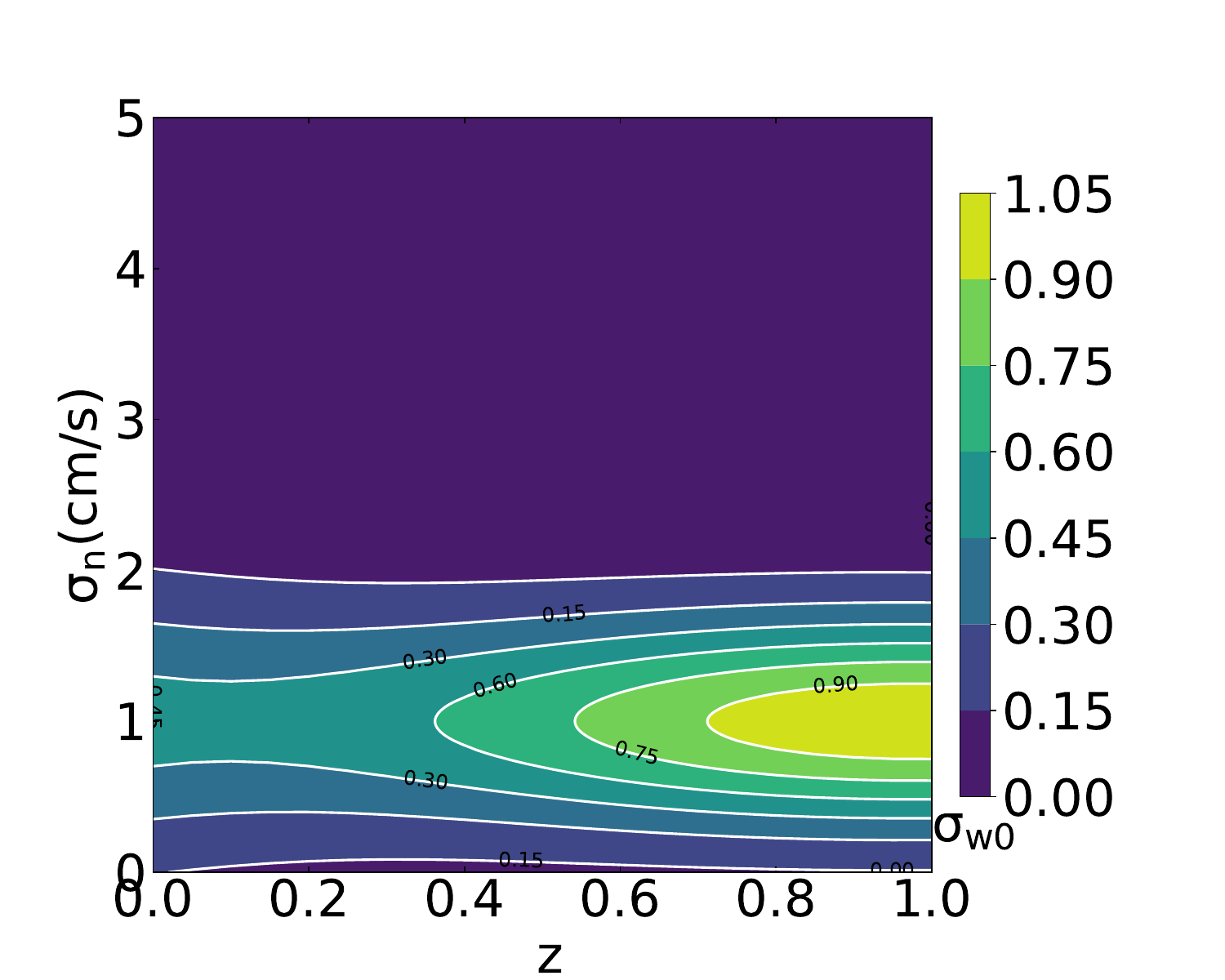}
  \includegraphics[width=.33\textwidth]{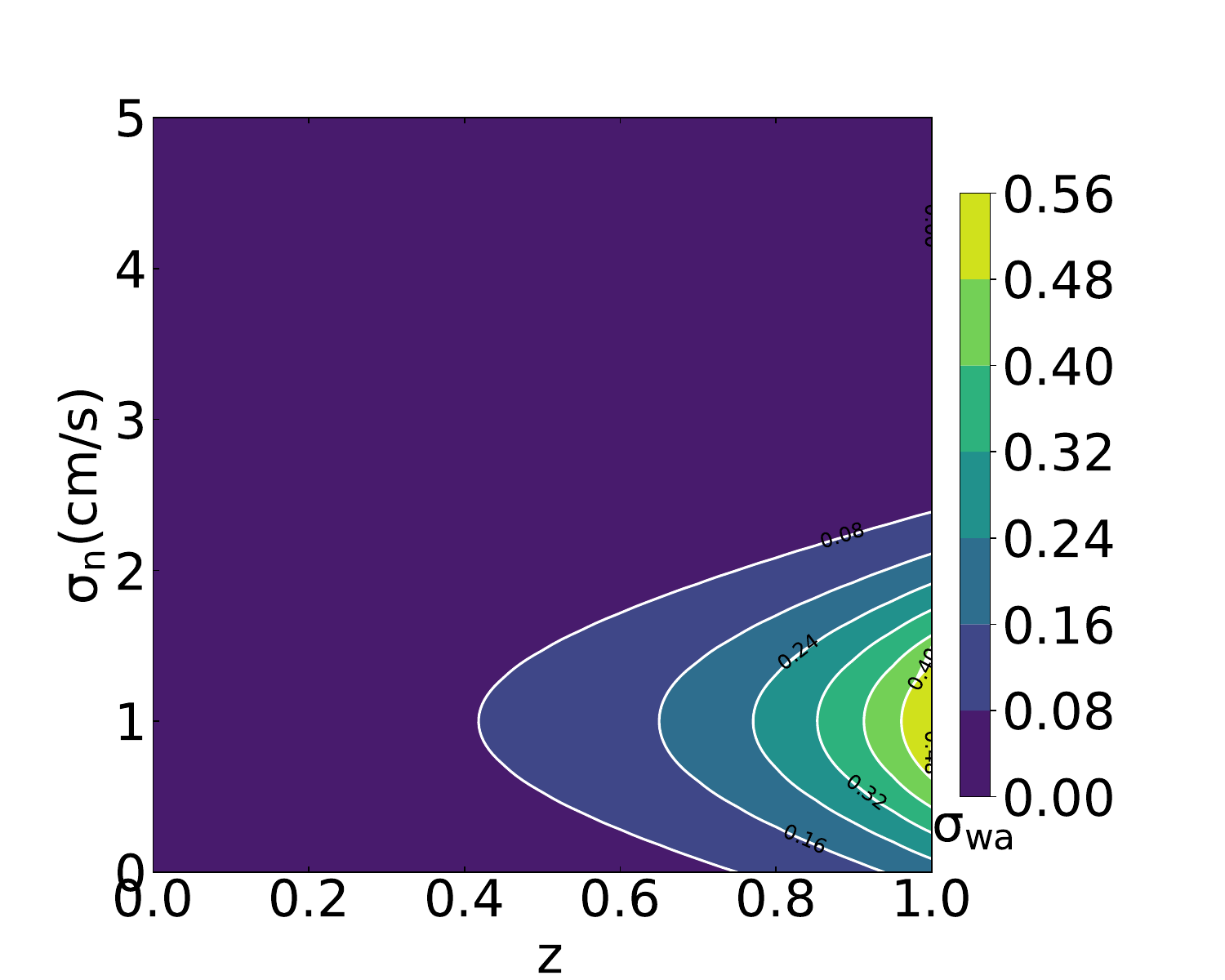}
\end{center}
\caption{The constraints on the precision of parameters $\sigma_m$, $\sigma_{w0}$, and $\sigma_{wa}$, derived from the spectral resolution data of 0.001 Hz over a 0.5-year observational period using the canonical redshift drift method, suggest that higher spectral resolution data will significantly enhance the precision of these constraints.}
\label{figs:sv2}
\end{figure*}

\begin{figure*}
	\begin{center}
		\includegraphics[width=.33\textwidth]{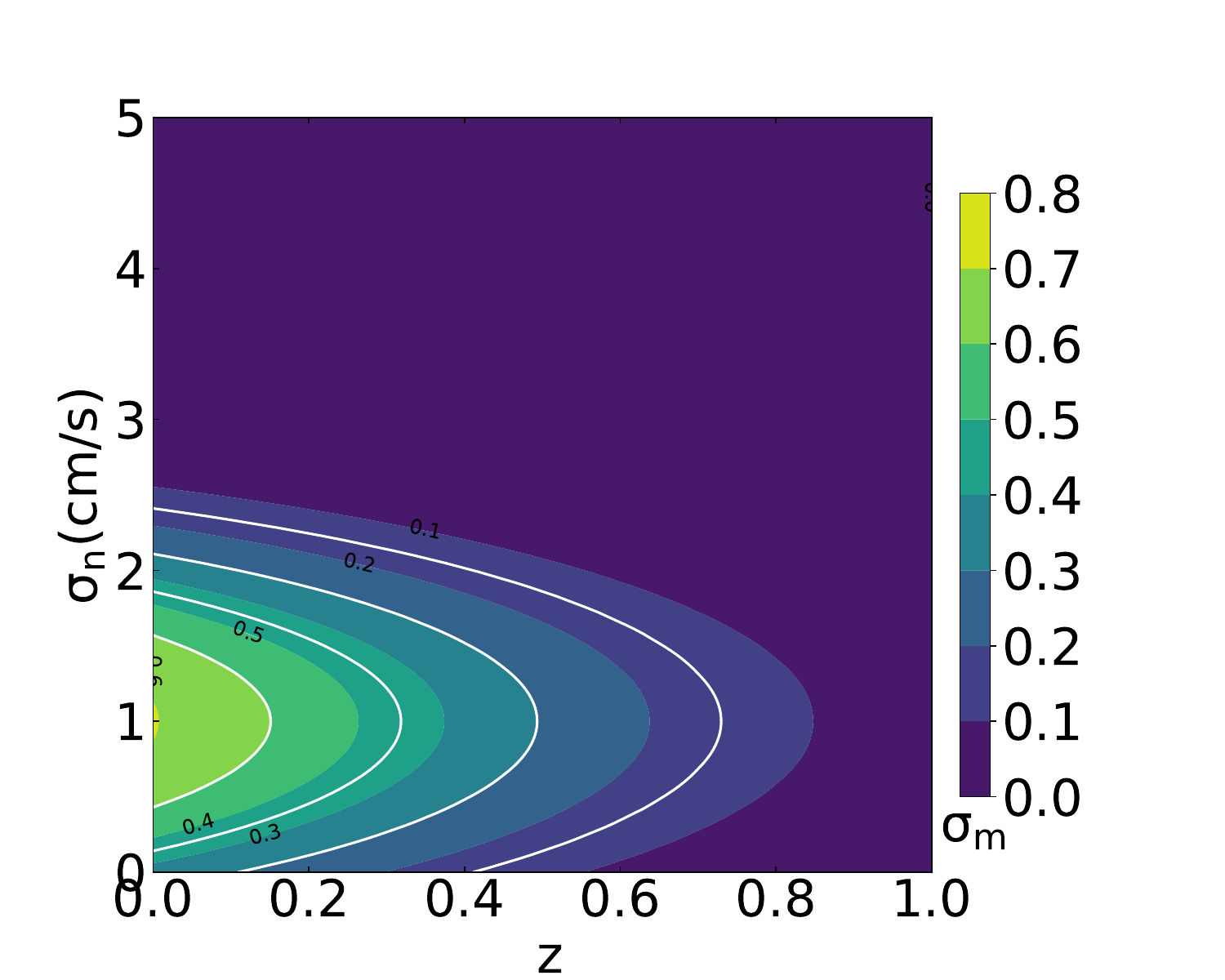}
		\includegraphics[width=.323\textwidth]{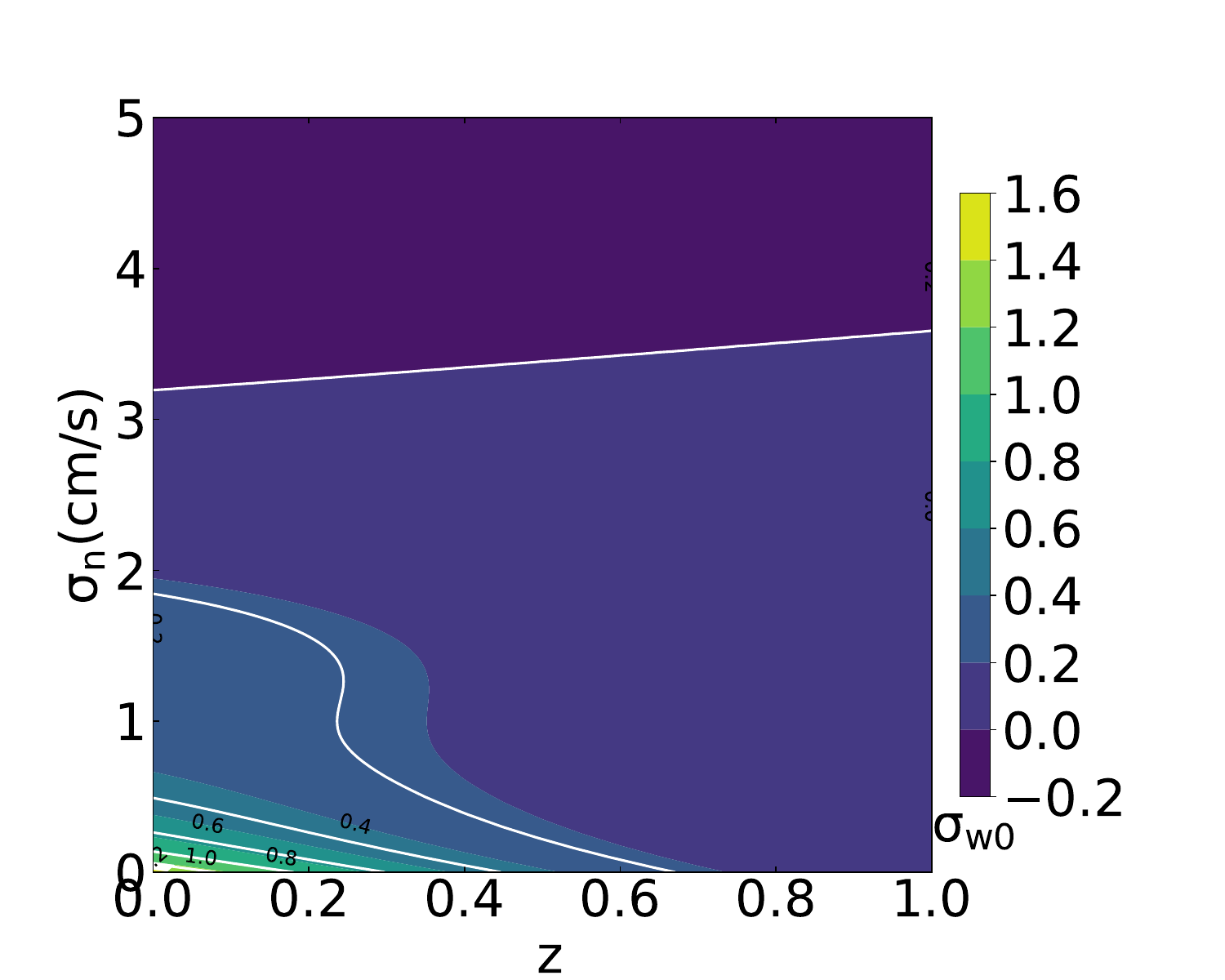}
		\includegraphics[width=.33\textwidth]{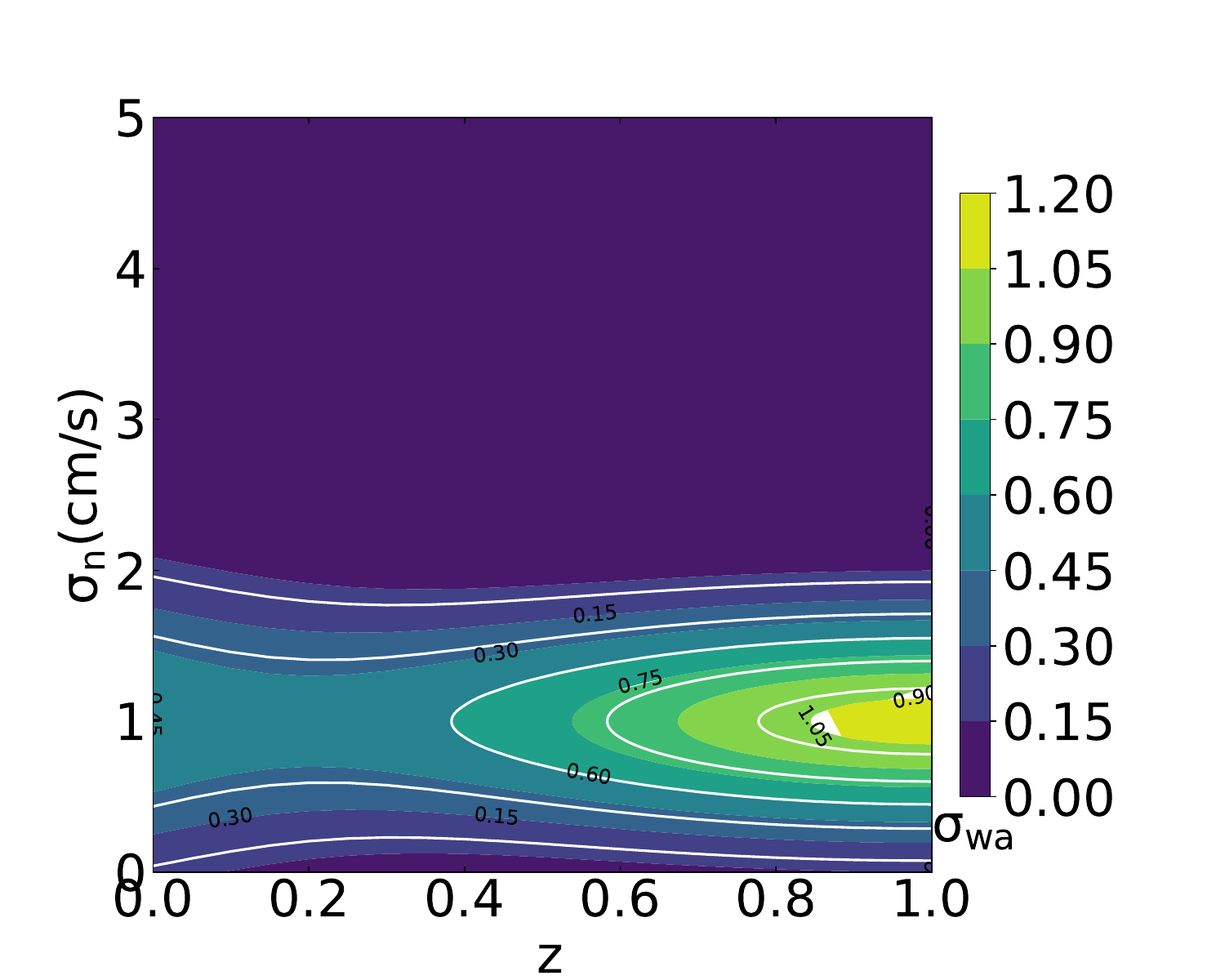}
	\end{center}
	\caption{The limitations on the precision of the parameters $\sigma_m$, $\sigma_{w0}$, and $\sigma_{wa}$ when employing spectral data with a frequency resolution of 0.002 Hz over an observational duration of $\Delta T = 0.5$ years using the standard redshift drift technique.}
	\label{figs:sv4}
\end{figure*}

The limitations on the precision of the parameter $\sigma_{p}$, specifically $\rm \partial (\Delta v)/\partial p$, where $p$ represents $\Omega_m$, $\omega_0$, and $\omega_a$, are derived from the CPL model framework, utilizing two datasets with spectral resolutions of 0.001Hz and 0.002Hz over an observational period of $\Delta T = 0.5$ year. The precision metrics for the cosmological parameters ($\sigma_m$, $\sigma_{w0}$, $\sigma_{wa}$), as indicated by the contours in figures \ref{figs:sv2} and \ref{figs:sv4}, reflect the constraints imposed by the two datasets based on the canonical redshift drift method. Table \ref{tab:CPL1} summarizes these results, with the 0.001Hz scenario in left-plot of figure \ref{figs:sv2} showing $\sigma_m$ ranging from 0.08 cm/s to 0.48 cm/s, $\sigma_{w0}$ in middle-plot exhibiting a broader range from 0.15 to 0.9 cm/s, and $\sigma_{wa}$ in right-plot falling between 0.08 to 0.4 cm/s. Conversely, the 0.002Hz data illustrated in figure \ref{figs:sv4} exhibits marginally relaxed values, with $\sigma_m$ spanning from 0.1 to 0.5 cm/s and $\sigma_{w0}$ demonstrating open contours exceeding 0.2 cm/s up to a redshift of 1. Furthermore, $\sigma_{wa}$ ranges approximately from 0.15 to 1.05 cm/s. These precision measurements substantiate that the two redshift drift datasets derived from spectral resolutions of 0.001Hz and 0.002Hz are both robust and superior to other observational data. Obviously, the higher spectral accuracy data enforces more stringent constraints compared to the lower resolution counterpart. On the whole, the precision for all parameters remains below the cm/s threshold over a period of $\Delta T$ = 0.5 year, consistent with theoretical predictions, whereas contemporary constraints on cosmological parameters from prevalent probes within this redshift range do not attain the precision of cm/s.

\begin{table}
	\centering
	\caption{The table lists the precision constraints for the parameters $\sigma_m$, $\sigma_{w0}$, and $\sigma_{wa}$, derived from spectral resolution data of 0.001 Hz and 0.002 Hz, over an experimental period of $\rm\Delta T$ = 0.5 year, utilizing the canonical redshift drift method. The results are illustrated in Figure \ref{figs:sv2} and Figure \ref{figs:sv4}, respectively.\label{tab:CPL1}}
	\begin{tabular}{l | c c | c c}
		\hline
		data & $\rm\sigma_m$(cm/s) & $\rm\sigma_{w0}$(cm/s) &  $\rm\sigma_{wa}$(cm/s)\\
		\hline
		0.001Hz & [0.08,0.48]& [0.15,0.9]&  [0.08,0.4]    \\
		0.002Hz &[0.1,0.5] & [0.2,1] & [0.15,1.05]  \\
		\hline
	\end{tabular}
\end{table}

\begin{figure*}
\begin{center}
\includegraphics[width=\columnwidth,keepaspectratio]{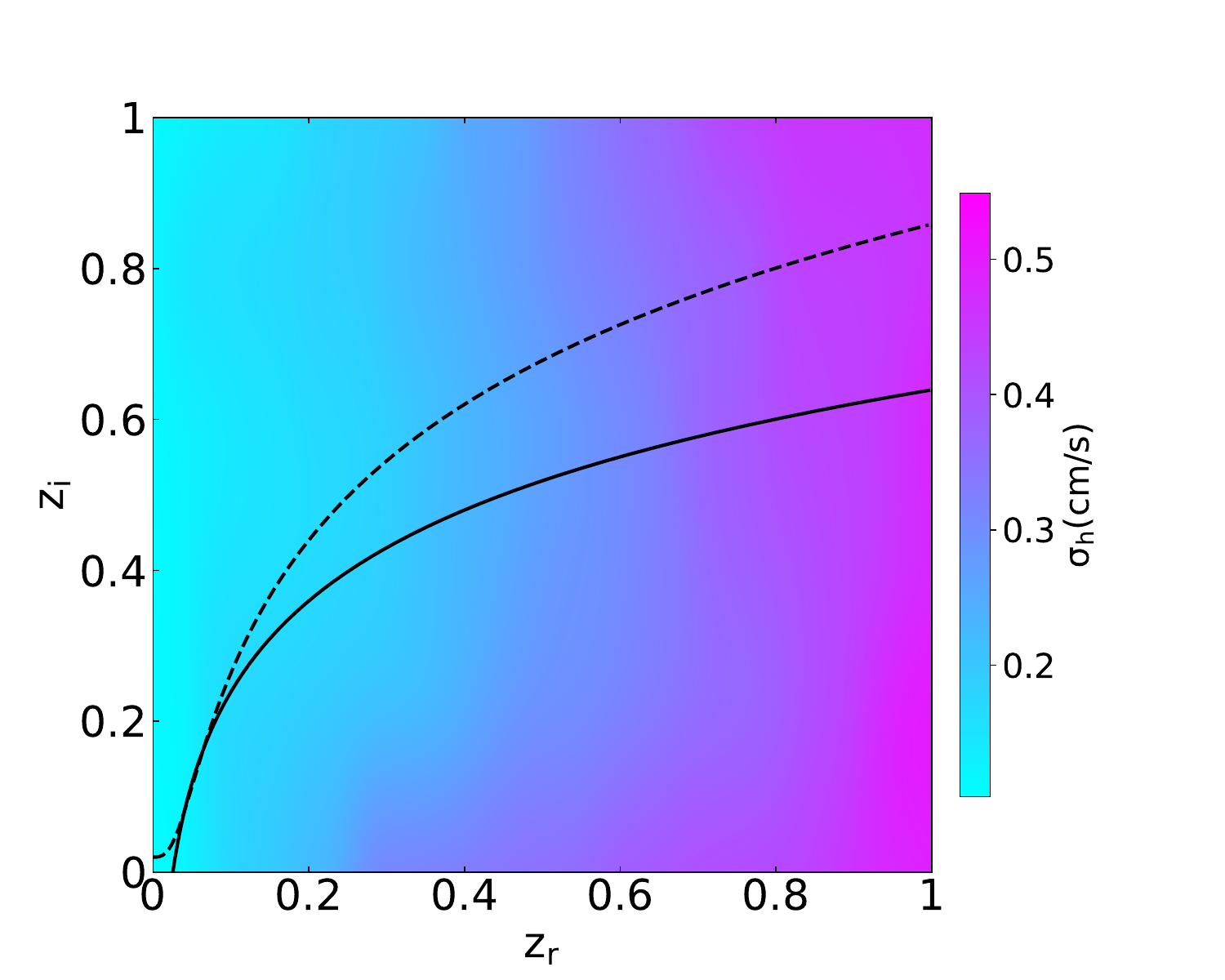}
\includegraphics[width=\columnwidth,keepaspectratio]{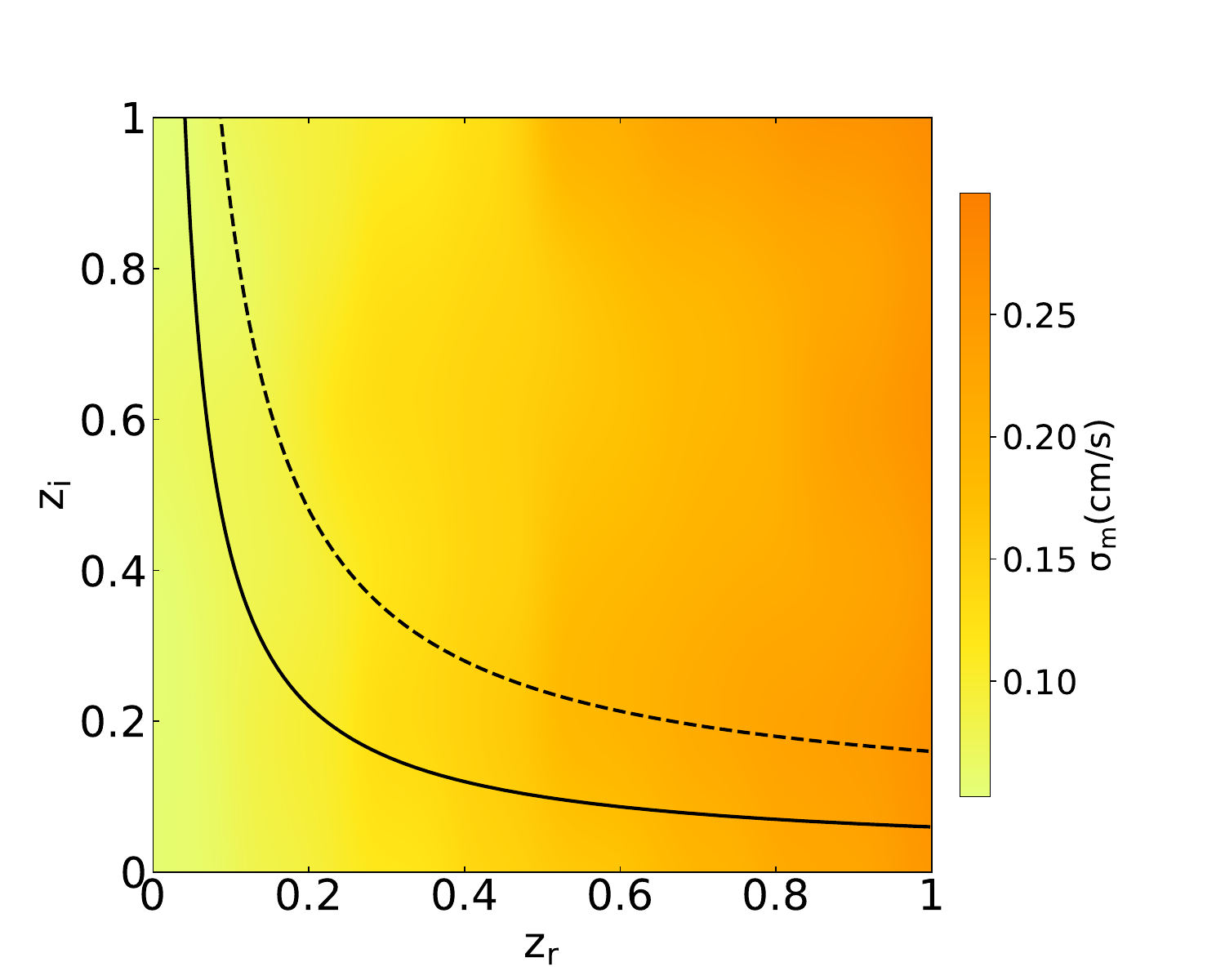}
\includegraphics[width=\columnwidth,keepaspectratio]{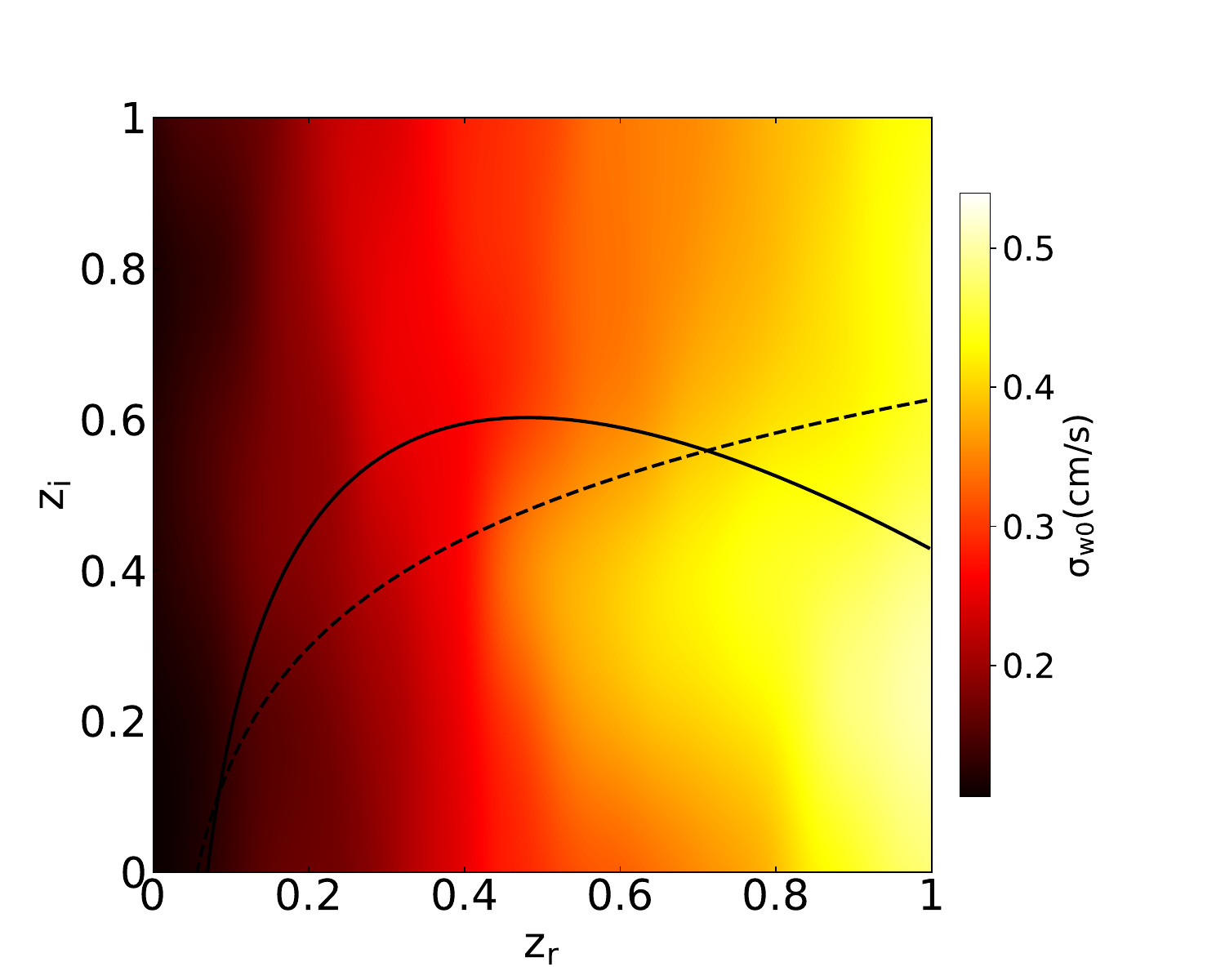}
\includegraphics[width=\columnwidth,keepaspectratio]{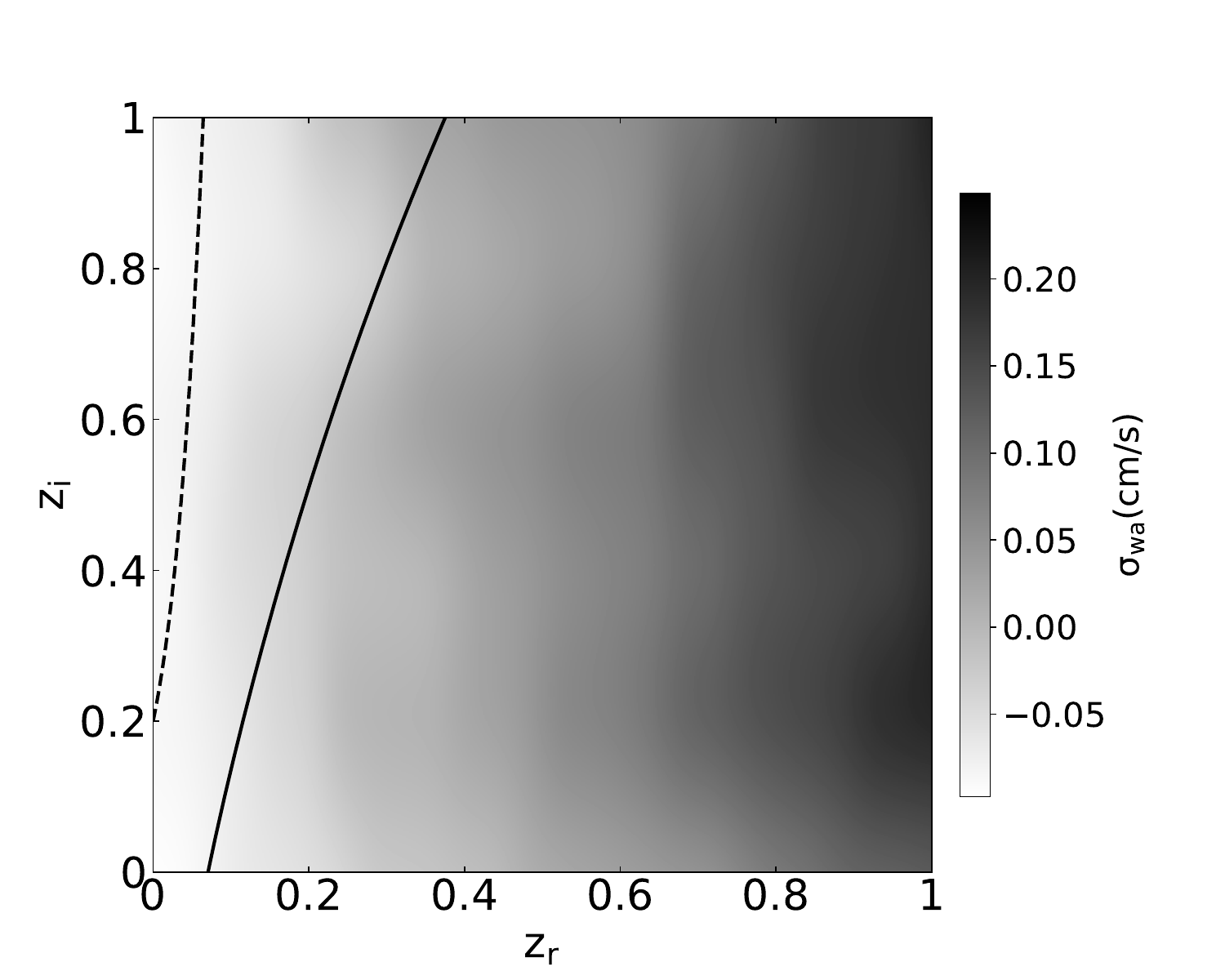}
\end{center}
\caption{The plotted differential redshift drift's  observational sensitivity  of cosmological parameters $h$ (top left), $\Omega_m$ (top right), $w_0$ (bottom left), and $w_a$ (bottom right) is depicted as a function of the reference redshift $z_r$ and the intervening redshift $z_i$. The colormap illustrates the sensitivity in cm/s for each 0.5-year observation period, with solid lines representing the 0.001 Hz data, and dashed lines indicating the 0.002 Hz scenario in each respective plot.}
\label{figs:ir}
\end{figure*}

For the differential redshift drift analysis, the sensitivity of parameters derived from redshift drift can be denoted as $\rm \partial  Sv / \partial {p_i}$ (where $p$ represents $h$, $\Omega_m$, $\omega_0$, and $\omega_a$) according to equation \ref{eqs:2}, \ref{eqs:s00}--\ref{eqs:s4}, considering a period of $\rm\Delta T = 0.5$ year and utilizing two datasets with frequency resolution of 0.001 Hz and 0.002 Hz. Figure \ref{figs:ir} depicts the precision  $\sigma_{p}$ for four parameters within the CPL model, plotted against the intervening redshift $\rm z_i$ and the reference redshift $\rm z_r$. The solid line corresponds to the constraint precision for the 0.001 Hz dataset, whereas the dashed line represents the 0.002 Hz dataset, with the color intensity indicating the measured precision values of these parameters in each plot. Additionally, the precision of each parameter generally declines as the redshift $\rm z_i$ or $\rm z_r$ surpasses 0.5. Specifically, for the panel of $\sigma_h$ in top-left in figure \ref{figs:ir}, which shows linear sensitivity to $\rm z_i$ and $\rm z_r$, the precision remains at approximately 0.1 cm/s of two datasets for redshifts below 0.1, subsequently, the precision for the 0.002 Hz dataset decreases beyond the values of the 0.001 Hz data, with the precision predominantly ranging between 0.1 and 0.5 cm/s.

In the plot of $\sigma_m$ in top-right, there is a more pronounced negative correlation between $\rm z_i$ and $\rm z_r$, resulting in tighter constraints. Notably, $\sigma_m$ demonstrates heightened sensitivity to the raised spectral resolution of the data, with a marked decline ranging from 0.01 to 0.25 cm/s due to the addition of the reference object's redshift $\rm z_i$. Moreover, the precision of 0.001 Hz data decreases more rapidly compared to 0.002 Hz data. In the bottom-left plot, $\sigma_{w0}$ exhibits similar behavior at lower redshifts for both datasets; however, it shows complexity and insensitivity to data resolution at median redshifts. Beyond a redshift of 0.7, the data becomes increasingly effective, aligning with anticipated precision trends. Regarding $\sigma_{wa}$ in the bottom-right plot, it is evident that both datasets are crucial for constraining $\sigma_{wa}$, resulting in a more narrowly defined precision range of -0.05 to 0.15 cm/s. The  corresponding measured values of four panels are summarized in the table \ref{tab:ir}.

\begin{table}
	\centering
	\caption{The table lists the comprehensive precision metrics of all parameters within the framework of differential redshift drift over an experimental duration of $\Delta T = 0.5$ year, as illustrated in figure \ref{figs:ir}. The continuous lines denote measurements at 0.001 Hz, whereas the dotted lines correspond to measurements at 0.002 Hz.}
	\label{tab:ir}
		\begin{tabular}{l | c c | c c}
			\hline
			data & 	$\sigma$(h) & 	$\sigma$($\Omega_m$) &$\sigma$($\omega_0$) & $\sigma$($\omega_a$)  \\
			\hline
			0.001Hz & [0.11,0.46]  & [0.04,0.26] & [0.1,0.47]  &[ -0.01,0.14] \\
			0.002Hz& [0.13,0.53]  &[0.08,0.3] & [0.11,0.48]  & [-0.01,0.2] \\
			 
			\hline
	\end{tabular}
\end{table}

The results definitively demonstrate that the two techniques, especially the differential redshift drift, exhibit superior precision in measurement compared to the traditional method, achieving an accuracy less than 0.5 cm/s with error margins reduced to 0.4 cm/s for each parameter derived from both data sets. Furthermore, the differential redshift drift significantly minimizes systematic measurement effects and lowers project costs for experiments conducted under similar conditions. Consequently, both the canonical method and the differential redshift drift provide more stringent constraints on the matter energy density $\rm\Omega_m$ and the dark energy equation of state parameters, $\rm w_0$ and $\rm w_a$, surpassing the precision of other probes, maintaining parameter accuracy within the range of below cm/s for spectral data sets of 0.001Hz and 0.002Hz. Specifically, for isolating dark energy constraints, the differential redshift drift is the more advantageous approach due to its lower uncertainties relative to the canonical method.
  
Overall, these empirical outcomes robustly corroborates both methodologies for detecting the redshift drift signal, particularly through the utilization of the HI 21 cm line that observed by SKA arrays. This study, however, concentrates on the lower redshift spectrum to examine the redshift drift phenomenon, with the objective of elucidating the characteristics of dark energy and quantitatively evaluating the rate of cosmic acceleration within the framework of real-time cosmology.

\section{summary}\label{sec:sum}

This study presents the precision constraints on cosmological parameters within the CPL model framework, utilizing redshift drift data derived from the detection of the HI 21cm signal by the SKA. The analysis leverages datasets with spectral resolutions of 0.001Hz and 0.002Hz collected over a 0.5-year observational period. The project demands dual observations over an extended time span, focusing on a distant astronomical object to trace the evolution of its redshift over cosmological timescales. The HI 21cm line is an ideal candidate for ground-based radio telescopes due to its immunity to atmospheric absorption and reflection. The redshift drift, characterized by its inherently subtle and purely radial nature, necessitates advanced techniques or specialized instrumentation for detection. It is anticipated that the SKA can demonstrate its exceptional capabilities in observational cosmology, particularly in exploring higher redshift domains, facilitated by its ultra-high-precision spectral data channels, thereby affirming the feasibility and robustness of this experimental investigation.

Two well-established methodologies exist for measuring this signal: the canonical redshift drift method, which is the standard method relative to the current epoch (redshift z=0), and the differential redshift drift technique, which utilizes two objects with non-zero redshifts, an intervening object redshift $z_i$ and a reference object redshift $z_r$. Figure \ref{figs:sd} illustrates the theoretical amplitude of both the canonical and differential redshift drift techniques as a function of reference redshift $\rm z_r$ over an observational period of $\Delta T$ = 0.5 year. It is evident that the differential redshift drift method facilitates a more detectable signal, owing to the marginally more pronounced magnitude of $\rm\Delta v$ compared to the canonical method. Integrating data from redshift drift observations with other cosmological probes can yield high-precision constraints on specific cosmological models or parameters, thus establishing a robust foundation for real-time cosmology.

Section \ref{sec:meas} describes the constrained results and firstly figure \ref{figs:sv1} demonstrates  that measured uncertainty of the spectroscopic velocity drift $\rm\sigma_{v}$ dependence with observatonal period $\Delta T$ and redshift z based on the equation  \ref{eqs:sv1}, suggesting the precision of error is significantly improved, achieving the level of below cm/s.  Conerning the constrainted precision of parameters in canonical method, the formalism of $\rm \partial (\Delta v)/\partial p$  is adopted to derive the precision in  data observed from the spectral resolution of 0.001 Hz and 0.002 Hz that shown in figure \ref{figs:sv2} -\ref{figs:sv4} and table \ref{tab:CPL1},  although the precision from  0.002Hz panel is slightly less stringent than that from 0.001Hz,  the precision still smaller the level of cm/s. By contrast, in case of the differential redshift drift approach, the method of $\rm \partial  Sv / \partial {p_i}$ is utilised to measure the sensitivty of parameters  that shown in figure \ref{figs:ir} and table \ref{tab:ir}, indicating the precision alway below 0.5 cm/s for any parameter. If the specific goal of constraining the matter energy density $\rm\Omega_m$, the observational strategy can be transformeded to the canonical redshift drift method, provided that the proper redshift of the targeted objects is selected. However, the differential redshift drift method highlights the distinct advantages when simultaneously constraining both the matter energy density $\Omega_m$ and the  parameters of equation of state of dark energy, $\rm w_0$ and $\rm w_a$.

In total, during the SKA epoch, the detection of the redshift drift signal, distinguished by its unparalleled spectral resolution, signifies a remarkably promising scientific endeavor through the HI 21 cm signal, with a specific emphasis on the absorption lines. The strategy involves observing numerous targets across various directions, ensuring sufficient exposure times to minimize systematic errors and elevate data quality. Adhering to these requirements will facilitate measurement precision to always be in the level of mm/s to cm/s or better, particularly emphasizing the differential redshift drift technique. This precision is pivotal for accurately ascertaining the cosmic acceleration rate in the context of real-time cosmology and for refining contemporary cosmological theories.
\section{Acknowledgements}
This work was supported by  National SKA Program of China, No.2022SKA0110202, National Key R\&D Program of China, No.2024YFA1611804, and China Manned Space Program through its Space Application System.

\section*{Data availability}
The data used in this article will be shared on reasonable request to
the corresponding author.
\bibliographystyle{mnras}
\bibliography{drift} 
 
\label{lastpage}
\end{document}